\documentclass{article}
\usepackage{spconf,amsmath,graphicx}
\usepackage{color}

\usepackage{import} 
\usepackage{algorithm}
\usepackage{algpseudocode}
\usepackage{makecell}
\usepackage[nolist]{acronym}
\usepackage{etoolbox}
\begin{acronym}
\acro{DNN}{Deep Neural Network}
\acro{DE}{Dialog Enhancement}
\acro{MAE}{Mean Absolute Error}
\acro{MSE}{Mean Squared Error}
\acro{BN}{Batch Normalization}
\acro{ANN}[ANN]{Artificial Neural Network}
\acro{SQC}{Sound Quality Component}
\acro{QEM}{Quality Estimation Module}
\acro{QM}{Quality Model}
\acro{PEM}{Parameter Estimation Module}
\acro{BSS}{Blind Source Separation}
\acro{BSSEval}{Blind Source Separation Evaluation}
\acro{SNR}{Signal to Noise Ratio}
\acro{SNRseg}{Segmental Signal to Noise Ratio}
\acro{fwSNRseg}{Frequency-Weighted Segmental Signal to Noise Ratio}
\acro{HAAQI}{Hearing-Aid Audio Quality Index}
\acro{HASQI}{Hearing-Aid Speech Quality Index}
\acro{HASPI}{Hearing-Aid Speech Perception Index}
\acro{LKR}{Log Kurtosis Ratio}
\acro{LLR}{Log-Likelihood Ratio}
\acro{dLLR}{Log-Likelihood Ratio Distance}
\acro{PESQ}{Perceptual Evaluation of Speech Quality}
\acro{PEAQ}{Perceptual Evaluation of Audio Quality}
\acro{POLQA}{Perceptual Objective Listening Quality Assessment}
\acro{PEASS}{Perceptual Evaluation methods for Audio Source Separation}
\acro{PEMO-Q}{PErception MOdel-based Quality}
\acro{PSM}{Perceptual Similarity Measure}
\acro{LPC}{Linear Prediction Coefficients}
\acro{CD}{Cepstral Distance}
\acro{SE}{Speech Enhancement}
\acro{SI}{Speech Intelligibility}
\acro{ASR}{Automatic Speech Recognition}
\acro{SDR}{Source to Distortion Ratio}
\acro{SIR}{Source to Interference Ratio}
\acro{SAR}{Source to Artifact Ratio}
\acro{SoNR}{Source to Noise Ratio }
\acro{ODG}{Overall Difference Grade}
\acro{MOS}{Mean Opinion Score}
\acro{MOV}{Model Output Variable}
\acro{NMR}{Noise to Mask Ratio}
\acro{ADB}{average distorted block}
\acro{EHS}{Harmonic Structure of the Error}
\acro{MFPD}{Maximum Filtered Probability of Detection}
\acro{WB}{Wide-Band}
\acro{OPS}{Overall Perceptual Score}
\acro{TPS}{Target-related Perceptual Score}
\acro{IPS}{Interference-related Perceptual Score}
\acro{APS}{Artifact-related Perceptual Score}
\acro{MNRU}{Modulated Noise Reference Unit}
\acro{STOI}{Short-Time Objective Intelligibility}
\acro{ITU-T}{International Telecommunication Union  - Telecommunication Standardization Sector}
\acro{ITU-R}{International Telecommunication Union - Radiocommunication Sector}
\acro{STFT}{Short-Time Fourier Transform}
\end{acronym}

\usepackage{psfrag}

\usepackage[pscoord]{eso-pic}% The zero point of the coordinate systemis the lower left corner of the page (the default).
\newcommand{\placetextbox}[3]{% \placetextbox{<horizontal pos>}{<vertical pos>}{<stuff>}
\setbox0=\hbox{#3}% Put <stuff> in a box
\AddToShipoutPictureFG*{% Add <stuff> to current page foreground
\put(\LenToUnit{#1\paperwidth},\LenToUnit{#2\paperheight}){\vtop{{\null}\makebox[0pt][c]{#3}}}%
}%
}%

\title{CONTROLLING THE PERCEIVED SOUND QUALITY FOR DIALOGUE ENHANCEMENT\\WITH DEEP LEARNING}

\makeatletter
\def\bstctlcite{\@ifnextchar[{\@bstctlcite}{\@bstctlcite[@auxout]}}
\def\@bstctlcite[#1]#2{\@bsphack
  \@for\@citeb:=#2\do{%
    \edef\@citeb{\expandafter\@firstofone\@citeb}%
    \if@filesw\immediate\write\csname #1\endcsname{\string\citation{\@citeb}}\fi}%
  \@esphack}
\makeatother

\renewenvironment{thebibliography}[1]{%
  \begin{oldthebibliography}{#1}%
    \setlength{\itemsep}{-.3ex}%
}%
{%
  \end{oldthebibliography}%
}

\name{Christian Uhle$^{\star \dagger}$ \qquad Matteo Torcoli$^{\star}$ \qquad Jouni Paulus$^{\star \dagger}$}
\address{ $^{\star}$ Fraunhofer Institute for Integrated Circuits IIS, Am Wolfsmantel 33, 91058 Erlangen, Germany \\$^{\dagger}$ International Audio Laboratories Erlangen\sthanks{A joint institution of the Friedrich-Alexander-University Erlangen-N\"urnberg (FAU) and Fraunhofer IIS, Germany.}, Am Wolfsmantel 33, 91058 Erlangen, Germany }

\begin{document}
\maketitle
\begin{sloppy}
\begin{abstract}
Speech enhancement attenuates interfering sounds in speech signals but may introduce artifacts that perceivably deteriorate the output signal.   
We propose a method for controlling the trade-off between the attenuation of the interfering background signal and the loss of sound quality. 
A deep neural network estimates the attenuation of the separated background signal such that the sound quality, quantified using the Artifact-related Perceptual Score, meets an adjustable target. 
Subjective evaluations indicate that consistent sound quality is obtained across various input signals. 
Our experiments show that the proposed method is able to control the trade-off with an accuracy that is adequate for real-world dialogue enhancement applications. 
\end{abstract}
\placetextbox{0.5}{0.08}{\fbox{\parbox{\dimexpr\textwidth-2\fboxsep-2\fboxrule\relax}{\footnotesize \centering Accepted paper, DOI: 10.1109/ICASSP40776.2020.9053789. \copyright  2020 IEEE. Personal use of this material is permitted. Permission from IEEE must be obtained for all other uses, in any current or future media, including reprinting/republishing this material for advertising or promotional purposes, creating new collective works, for resale or redistribution to servers or lists, or reuse of any copyrighted component of this work in other works.}}}%
\bstctlcite{IEEEexample:BSTcontrol}
\acresetall
\section{Introduction}
\label{sec:intro}
Speech enhancement processes the input signal \mbox{$x(n) = s(n) + b(n)$} with the aim to improve the intelligibility of the speech signal $s(n)$ by attenuating the interfering background signal $b(n)$. 
This can be applied for dialogue enhancement in TV and movie sound when the level of the speech is too low compared to the level of environmental sounds and music in the background~\cite{Herre14-JAES, Fuchs14-SMPTE, Shirley15-JAES}. 
The processing may introduce artifacts that deteriorate the perceived sound quality.

Data-driven methods, e.g.,~using \acp{ANN} \cite{Tchorz2003, Kleinschmidt2003, Uhle08-AES}, estimate a representation of the target signal or the parameters for retrieving the target signal from the input mixture.  
While most methods optimize a cost function without taking perceptual constraints into account, perceptually motivated cost functions have been developed based on \ac{STOI} \cite{fu:2018, koizumi:2018, zhao:2018}, \ac{PESQ}, and \ac{PEASS}~\cite{koizumi:2017}. 
Yet, no control of the trade-off between sound quality and attenuation is possible.  

Two related works apply \acp{DNN} to predict the sound quality of \ac{BSS} quantified by means of \ac{SDR} (which is then used to select the best sounding output signal from multiple \ac{BSS} methods)~\cite{manilow:2017} and \ac{SAR}~\cite{grais:2018}. 
These methods do not facilitate the control of sound quality,  and the measures \ac{SDR} and \ac{SAR}~\cite{Vincent:2006} were shown to correlate poorly with the perception of  sound quality~\cite{torcoli:2018, cano:2016, cartwright:2016}.

Here, we present a single-ended method for controlling the trade-off between background attenuation and sound quality for speech enhancement.  
We propose to attenuate the background only partially such that the sound quality of the output meets a target level. 
To this end, a \ac{DNN} is trained to estimate the background attenuation parameter with target values that are obtained from a computational model of sound quality. % for which we use the \ac{APS}. 

The proposed method is evaluated with signals that are representative for the application of dialogue enhancement. 
To the best of our knowledge this is the first method for controlling the perceived sound quality for speech enhancement applications. 
The paper is structured as follows: Section \ref{sec:control} details the proposed method, experimental results are given in Section \ref{sec:eval}, and Section \ref{sec:conclusion} concludes the paper.

\begin{figure*}[th]
\centering
\def\svgwidth{0.75\textwidth}
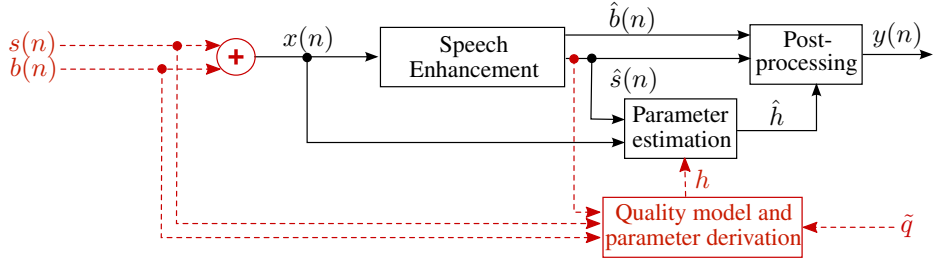
  \caption{\label{fig:postpro}System overview of the proposed method. The processing shown in solid black lines is applied for dialogue enhancement. Red dashed lines illustrate the processing that is only required for the off-line training of the parameter estimation.}
\end{figure*} 

\section{Proposed control of sound quality}
\label{sec:control}
We assume that the degradation of the sound quality increases monotonically with the attenuation of the background signal. 
Our aim is to adjust the background attenuation such that the perceived sound quality equals a target value $\tilde q$. 
The sound quality level $\tilde q$ determines the desired trade-off between separation and sound quality and is adjustable to meet the needs of the user in their listening environment. 
The input signal is decomposed into estimates $\hat s(n)$ and $\hat b(n)$ of the target and the background signal. 
The output signal $y(n)$ is computed as  
\begin{equation}	
	y(n)= \hat s(n) + g  \hat b(n),
	\label{eq:drywet}
\end{equation}
where the background attenuation $g$ is computed using supervised regression trained with target values from a computational model for sound quality. 
Fig.\,\ref{fig:postpro} shows an overview of the proposed method. 

\begin{table*}
\begin{footnotesize} % footnotesize
\begin{center}
\begin{tabular}{|c|c|c|c|c|c|c|c|c|c|c|}
\hline
{\textbf{Layer}}
&{\textbf{In}}
&{\textbf{Conv2D}}
&{\textbf{MaxP2D}}
&{\textbf{Conv2D}}
&{\textbf{MaxP2D}}
&{\textbf{Conv2D}}
&{\textbf{MaxP2D}}
&{\textbf{Flat.}}
&{\textbf{Dense}}
&{\textbf{Dense Out}} \tabularnewline
%&{\makecell{\textbf{Dense} \\ \textbf{Out}}} \tabularnewline
\hline
% ------------------------------------
\makecell{\# Units \\ / Filters }
& \makecell{STFT coefficients}
& 32
& --
& 64
& --
& 128
& --
& --
& 256
& 1 \tabularnewline
\hline 
% ------------------------------------
\makecell{Output \\ Shape}
& 2, 374, 257
& 32, 374, 257
& 32, 94, 65
& 64, 94, 65
& 64, 24, 17
& 128, 12, 9
& 128, 6, 5
& 3840
& 256
& 1 \tabularnewline
\hline 
% ------------------------------------
\makecell{Filter Size \\ / Stride}
& --
& \makecell{16, 16 \\ 1, 1}
& \makecell{8, 8 \\ 4, 4}
& \makecell{8, 8 \\ 1, 1}
& \makecell{8, 8 \\ 4, 4}
& \makecell{4, 4 \\ 2, 2}
& \makecell{4, 4 \\ 2, 2}
& --
& --
& -- \tabularnewline
\hline 
% ------------------------------------
Activation
& --
& ReLU
& --
& ReLU
& --
& ReLU
& --
& --
& ReLU
& ReLU \tabularnewline
\hline 
% ------------------------------------
Notes
& \makecell{ 2 signals x \\ 4\,s audio \\ at 12\,kHz }
& \makecell{L2 Reg \\ 0.001}
& \makecell{Padding \\ same}
& \makecell{L2 Reg \\ 0.001}
& \makecell{Padding \\ same}
& \makecell{L2 Reg \\ 0.001}
& \makecell{Padding \\ same}
& --
& \makecell{Dropout \\ 30\%}
& \makecell{Control \\ param.\\ (dB)} \tabularnewline
\hline 
% ------------------------------------
\# Param.
& --
& 16,416        
& 0
& 131,136    
& 0
& 131,200    
& 0
& 0
& 983,296    
& 257 \tabularnewline
\hline 
% ------------------------------------
\end{tabular}
\end{center}
\end{footnotesize}
\vspace{-10pt}
\caption{\label{tab:arch}Information on structure and parameters of the DNN for the parameter estimation.}
\end{table*}

\subsection{Speech enhancement implementation}
\label{sec:separation}
The speech enhancement method from~\cite{paulus:2019} is used in our experiments to compute training and test signals for the proposed control method. 
It has been extensively tested for dialogue enhancement of archived broadcast material in the context of object-based audio~\cite{torcoliAST:2018}. 
The method applies real-valued weights to the \ac{STFT} representation of the input signal computed for 21.3\,ms frames with 50\% overlap.   
It uses a combination of center signal extraction, primary-ambient decomposition, semi-supervised non-negative matrix factorization with a spectral basis dictionary for speech~\cite{Kim15-SPL}, and single-channel speech enhancement using an iterative level estimation for stationary noise signals~\cite{ephraim:1984, ephraim:1985}. 
The spectral weights from these methods are combined using element-wise minimum operation and are signal-adaptively smoothed to reduce musical noise~\cite{esch2009}. 
The processing reduces the \ac{SIR} by 9.4 and 13.1\,dB for mono and stereo signals, respectively, on average over the test data set used in~\cite{paulus:2019}.   

\subsection{Computational model for sound quality}
The sound quality is computed by means of the \ac{APS} with the implementation from the \ac{PEASS} toolbox~\cite{emiya:2011}.
\ac{APS} predicts the outcome of a MUSHRA test  \cite{ITUBS1534_2003}, where  listeners would rate the quality of the signal in terms of absence of additional artificial noise. 
The input signal is decomposed with orthogonal projections in the Gammatone domain into signal component representing the target, the interferer, and artifacts.
\mbox{PEMO-Q}~\cite{PEMOQ:2006} is used to compute the input features for an \ac{ANN} that is applied to predict the \ac{APS} score and three other quality measures.

\subsection{Parameter estimation}
\label{sec:parameter_estimation}
We train a \ac{DNN} to estimate the background attenuation in dB, $h = -20 \log_{10} (g)$, with $0.01\leq g \leq 1$, such that the output sound quality meets a target level.  
The inputs to the \ac{DNN} are the separated speech signal $\hat s(n)$ and the input mixture \mbox{$x(n)$}. 
The target values $h$ for the training are computed with an iterative line search, where the update direction and step size are determined based on the error in the resulting quality level as
\begin{equation}	
	h_{k+1} = h_k - \gamma (\tilde q - q), 
	\label{eq:update}
\end{equation}
with iteration index $k$, \ac{APS} value $q$ for $h_k$, target quality level $\tilde q = 80$, step size $\gamma=0.5$ and $h_0=20$. 
The update is repeated for 6\,iterations at most or until the stop criterion ${|q-\tilde q|<0.25}$ has been reached.  
Items for which \mbox{$|q-\tilde q| \geq 1$} are discarded. 
The target values $h$ and the \ac{DNN} outputs $\hat h$ are computed for non-overlapping signal segments of 4\,s length each. 

\subsection{Structure and training of the DNN}
The audio signals are downsampled to 12\,kHz for reducing the computational complexity and the number of network parameters.  
The log-magnitude \ac{STFT} coefficients are computed using a sine window of size 256\,samples, 50\% overlap, and transform length of 512. 
The input to the \ac{DNN} is a 3D tensor with shape 2 (unprocessed signal and estimated speech) $\times$ 374 (time frames) $\times$ 257 (frequency bins). 
We centered and normalized the data using means and standard deviations computed from the training data along the time axis.  

Table~\ref{tab:arch} shows the structure and hyperparameters of the network. 
It uses three series of a 2D convolutional layer, a max-pooling layer, and a \ac{BN} layer. 
They are followed by a first dense layer, a \ac{BN} layer, and a final dense layer.  
The \ac{DNN} is trained by minimizing the \ac{MSE} using mini-batch gradient descent with a batch size of 64, momentum of 0.5, and Nesterov acceleration.  
The training is done for 60 epochs with a learning rate of $10^{-5}$, followed by three additional epochs of refinement with a learning rate of $10^{-6}$. 

\section{Evaluations}
\label{sec:eval}
\subsection{Training and testing data}
The training data are created from 13\,hours of mixtures of clean speech with various background signals. 
The speech signals feature female and male talkers, various languages, accents, and speech rates. 
Approximately half of the background signals are environmental noise and sound effects that originate from four libraries of sound effects for audio productions. 
The other half is instrumental music sampled from commercial recordings. 
The initial data is augmented by a factor of 5 by mixing speech and background signals at \acp{SNR} of $[-10, 0, 5, 10, 20]\,\mbox{dB}$.

Fig.\,\ref{fig:data} depicts the distribution of the target values $h$ for the training data together with a weighting function. 
The distribution is unimodal with a maximum at about 14\,dB.
In order to avoid biased estimation towards this value, the weighting function is applied to the loss values.

\begin{figure}[]
\centering
\def\svgwidth{0.98\columnwidth}
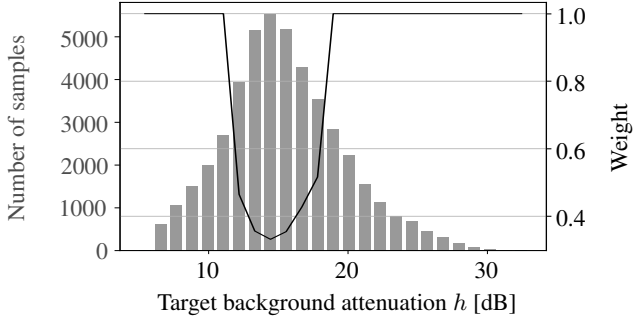
  \caption{\label{fig:data}Distribution of the target values in the training data (bar plots) and training weights (solid line). }
\end{figure} 

The test data comprise 180\,items with a length of 4\,s each that have been created by mixing 36 different speech and background signals at the same 5 different \acp{SNR}.  

\begin{figure}[]
\centering
\def\svgwidth{0.98\columnwidth}
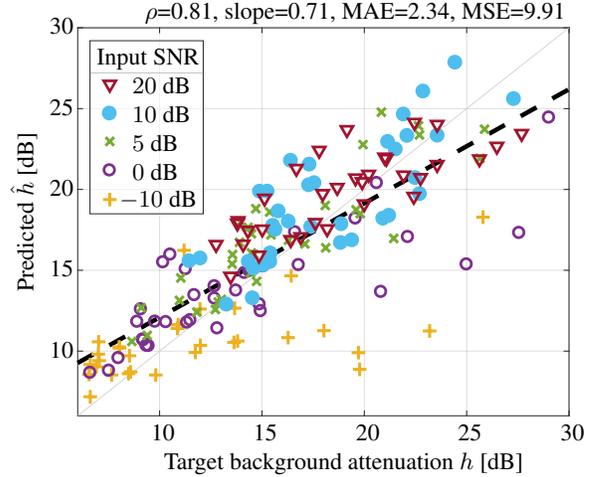
\caption{\label{fig:results}Predictions over input \acp{SNR} and regression function.}
\end{figure} 

\subsection{Objective evaluation results}
The \ac{DNN} estimates the background attenuation $h$ with a \ac{MAE} of 2.34\,dB and an \ac{MSE} of 9.91\,dB.
Fig.\,\ref{fig:results} shows the target and predicted values for each item of the test set, which correlate with a $\rho$ coefficient of 0.81 (p-value=0.00). 
The linear regression between target and predicted values has a slope of 0.71 with an r-squared value of 0.65.
 
Fig.\,\ref{fig:bars} shows the \ac{MAE}, averaged over all input \acp{SNR} for each test item (left plot) and averaged over all test items for each input \ac{SNR} (right plot). 
We can see that the performance varies largely among the test items and that both, the \ac{MAE} and its variance, are largest for the lowest \acp{SNR}.
A detailed inspection of the worst performing item revealed that the bad overall performance is caused by the two lowest \acp{SNR}.

The \ac{APS} of the output test signals when mixed with $\hat h$ has a mean of 82.6 and a standard deviation of 4.50. 
The \ac{MAE} between the output \acp{APS} and the target APS of 80 is 4.48. 

 \begin{figure}[hb]
\centering
\def\svgwidth{0.95\columnwidth}
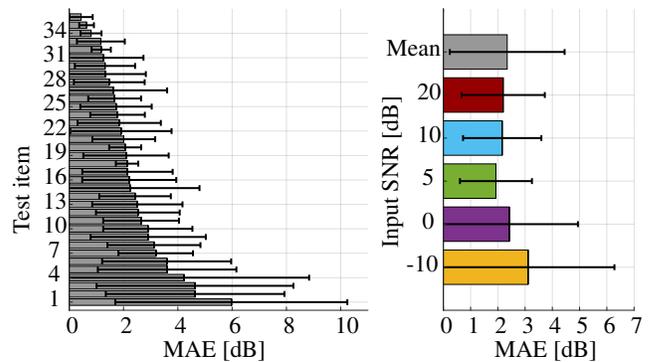
\caption{\label{fig:bars}\acf{MAE} (depicted by bars) and standard deviation (overlapping lines) for the test items over the different input \acp{SNR} (left plot) and for the different input \acp{SNR} over the test items (right plot).}
\end{figure} 

\subsection{Modification and ablation analysis}
The structure of the \ac{DNN} is inspired by~\cite{manilow:2017} with the following modifications.  
We reduced the number of dense layers from five to two which reduced the number of parameters by about \,50,000 and improved the \ac{MSE} by about 50\%. 
The \ac{BN} layers are placed after each pooling layer and before the last dense layer, while in~\cite{manilow:2017} the \ac{BN} layers are positioned before the pooling layers and they are not used between the dense layers. 
Moving the \ac{BN} layers to before the pooling layer caused a performance degradation in \ac{MSE} of ca.\,10\%.

Further modifications from~\cite{manilow:2017} are the use of log-magnitude \ac{STFT} coefficients instead of time signals as inputs to the \ac{DNN}, and differences in the convolution filter dimensions and the activation functions. 
All modifications have led to lower regression errors in our experiments. 

When the weights in Fig.\,\ref{fig:data} were not used during training, almost the same $\rho$, \ac{MAE}, and \ac{MSE} are obtained on the test data, but the regression slope would be 0.60 (r-squared=0.64). 
Thus, using sample weighting while training causes slightly worse predictions for values close to 14\,dB, but better predictions for less frequent values in the training data. 
An additional experiment showed that if only the estimated speech signal is used as input to the network (without the signal before separation), the \ac{MSE} increases by about 50\%.

\subsection{Subjective evaluation results}
The proposed method has been evaluated in a listening test with 13 listeners without reported hearing impairments. 
Eight input signals were created by mixing speech with background sounds and processed with  $h=\hat h$.  
We tested two additional conditions with background attenuations of $\hat h-6\,\mbox{dB}$ and $\hat h+6\,\mbox{dB}$.
The items were presented one at a time in random order and the listeners were asked to rate the absolute sound quality in terms of absence of artifacts or distortions without a reference. The discrete 5-point annoyance scale \cite{ITUTP800:1996} (1=very annoying, 2=annoying, 3=slightly annoying, 4=audible but not annoying,  5=inaudible) was employed. 
 
\begin{figure}[t]
\centering
\def\svgwidth{1\columnwidth}
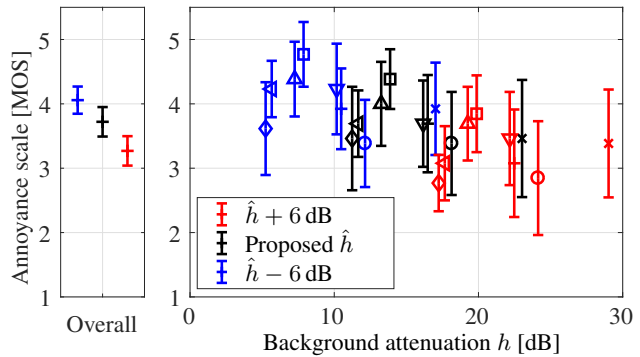
\caption{\label{fig:lt} Mean and 95 \% confidence intervals of the listening test results over all listeners and items (left plot) and per item as function of the applied background attenuation $h$ (right plot). In the right plot, the same symbol is used for the items originating from the same input mixture.}
\end{figure}  
 
Fig.\,\ref{fig:lt} shows the Mean Opinion Scores (MOS) and confidence intervals over all listeners and items (left plot) and per item as function of the applied background attenuation $h$ (right plot). 
Although $11.2\,\mbox{dB} \leq \hat h \leq 23 \,\mbox{dB}$ spans a wide range for the different input signals, all MOS for the proposed condition are consistent within a small range of  $[3.38, 4.38]$ and close to being ``audible but not annoying'' on average.  

\section{Conclusion}
\label{sec:conclusion}
This paper proposed a method for controlling the output sound quality of speech enhancement by adjusting the attenuation of the interfering background signal such that the perceived quality meets a target level. 
To this end, we trained a \ac{DNN} with target values obtained from \ac{APS} to predict the background attenuation for the speech enhancement method described in \cite{paulus:2019}. 
For optimum results, we suggest to re-train the parameter estimation for other speech enhancement methods if the introduced artifacts have different characteristics.

The background attenuation is estimated with an \ac{MAE} of 2.34\,dB, resulting in an \ac{MAE} of 4.48 for the \ac{APS} obtained for the output signals of our test set. 
With the chosen target quality level of $\tilde q = 80$ \ac{APS} points the proposed method achieved a mean score slightly below ``audible but not annoying'' in a listening test.  
While the target quality level appears to be a good choice for hearing impaired listeners, larger values may be adequate for other target groups.

The calculation of \ac{APS} is computationally expensive and requires reference signals for the clean speech which are not available in the application. 
The advantage of the proposed method is that the \ac{APS} is only required during training. 
Future work will assess the system performance by means of a listening test and investigate the relation between target quality for the parameter estimation and perceived sound quality. 

\bibliographystyle{IEEEtran}
\bibliography{Bibliography_long}

\end{sloppy}
\end{document}